\documentclass[journal]{IEEEtran}

\usepackage{url}
\usepackage{graphicx}
\graphicspath{ {./} }
\usepackage{tablefootnote}
\usepackage{array}
\usepackage{multirow}
\usepackage{amsmath}
\usepackage{adjustbox}
\usepackage{booktabs}
\usepackage{makecell}

\usepackage{algorithm}
\usepackage{algorithmicx}
\usepackage{algpseudocode}

\usepackage[most]{tcolorbox}

\definecolor{background}{rgb}{0.98, 0.98, 0.98}
\definecolor{comment}{rgb}{0.3,0.6,0.3}
\definecolor{keyword}{rgb}{0.8,0.1,0.1}
\definecolor{string}{rgb}{0.5,0.1,0.7}

\usepackage{listings}
\usepackage{xcolor}

\lstdefinestyle{promptstyle}{
  backgroundcolor=\color{gray!3},
  basicstyle=\ttfamily\scriptsize,
  frame=single,
  rulecolor=\color{gray!10},
  breaklines=true
  framerule=0pt,
  xleftmargin=2pt,
  xrightmargin=8pt,
  aboveskip=10pt,
  belowskip=0pt,
  captionpos=b,
  showstringspaces=false,
  morekeywords={Role,Input,Instructions,Output}
}

\lstdefinestyle{promptstyle*}{
  backgroundcolor=\color{gray!5},
  basicstyle=\ttfamily\scriptsize,
  frame=single,
  framerule=0.5pt,
  rulecolor=\color{gray!5},
  numbers=left,
  numberstyle=\tiny\color{white},
  breaklines=true,
  showstringspaces=false,
  tabsize=1,
  captionpos=b,
  morekeywords={Role,Input,Instructions,Output}
}

\begin{document}
%
\title{ASAD: Adaptive Software Agents for Debugging}
%
%
%

\author{
    Yacine~Majdoub, Eya~Ben~Charrada, and Haifa~Touati%
    \thanks{All authors are with the IReSCoMath Research Laboratory, Faculty of Sciences, University of Gabes, Tunisia 
    (emails: yacine.majdoub@enig.rnu.tn, eya.bencharrada@fsg.rnu.tn, haifa.touati@univgb.tn).}%
    \thanks{Yacine Majdoub is also with the National School of Engineering, Gabes, Tunisia.}%
}

%
%

\markboth{IEEE TRANSACTIONS ON SOFTWARE ENGINEERING}%
{Majdoub \MakeLowercase{\textit{et al.}}: Understanding and Mitigating Library-Related
Issues in LLM-Generated Code}
%



\maketitle

\begin{abstract}
The integration of Large Language Models (LLMs) into multi-agent systems has shown great potential for automated debugging. Yet nearly all current frameworks rely on rigid, predefined architectures: the number of agents, their roles, and their interaction patterns are fixed before any analysis of the bug occurs. This one-size-fits-all approach is fundamentally mismatched to the heterogeneous nature of software defects. Simple bugs waste resources on unnecessary coordination, while complex ones suffer from insufficient or poorly aligned expertise.

This paper introduces ASAD, an adaptive agentic system for debugging that configures its team according to the nature and complexity of each bug. ASAD initiates the debugging process by analyzing the faulty code and dynamically determines the number of agents to deploy, the specialized roles they should have, and the collaboration strategy they should follow. A central coordinator orchestrates this process through iterative planning, reflection, and execution; applying fast single-pass repairs for simple issues while assembling purpose-built teams to tackle more complex failures.
\textcolor{black}{We evaluate ASAD on three established benchmarks: Defects4J, DebugBench, and CodeFlaws, using multiple LLMs, such as DeepSeek-V3, \textcolor{black}{Qwen-3} and GPT-5. ASAD consistently improves bug-fix rates by 12–20\% over chain-of-thought(CoT) prompting and consistently outperforms static multi-agent systems by 4–9\% in fix precision while reducing average agent usage by 32\%.} Crucially, our system dynamically adjusts the number and roles of agents: it resolves simple bugs with minimal coordination and scales agent involvement only for more complex cases.

\end{abstract}

\begin{IEEEkeywords}
Adaptive LLM, Agentic AI Systems, Software Debugging
\end{IEEEkeywords}

%
\IEEEpeerreviewmaketitle

\section{Introduction}


Large Language Models (LLMs) have fundamentally reshaped software engineering research and practice. Their capacity to interpret natural language, reason over code semantics, and generate structured outputs has driven significant advances in code generation~\cite{joel2025surveyllmbasedcodegeneration}, automated bug repair~\cite{Majdoub2024Debugging}, and test synthesis~\cite{wang2024softwaretestinglargelanguage}. These capabilities are further enhanced by reasoning strategies, such as task decomposition, self-reflection, and iterative refinement, that enable LLMs to tackle increasingly complex tasks~\cite{agenticDesignP}.

A growing body of work explores multi-agent systems, where multiple LLM-powered agents collaborate with specialized roles to solve software engineering problems~\cite{han2024llmmultiagentsystemschallenges}. In debugging, for instance, distinct agents may handle fault localization, patch synthesis, static analysis, and validation~\cite{lee2024unifieddebuggingapproachllmbased}. This division of responsibilities mimics human software teams, where collaboration and specialization lead to better results. Such architectures have been shown to improve reasoning quality, reduce hallucinations, and increase success rates in program repair~\cite{tang2026llmbasedagenticsystemssoftware}.

Despite this promise, current multi-agent debugging systems suffer from a critical design flaw. Most existing approaches rely on static and rigid architectures in which the number of agents, their roles, and their communication patterns are predefined and remain unchanged throughout execution. ~\cite{he2024llmbasedmultiagentsystemssoftware}.
This lack of adaptability limits the scalability, the same configuration of agents is applied regardless of the complexity of problems, leading to inefficient use of computational resources and avoidable costs in time and computation, particularly for cases that could be effectively handled by a single agent, or suffering from insufficient or misaligned expertise for complex tasks ~\cite{han2024llmmultiagentsystemschallenges}.
Moreover, prolonged interactions among agents often reduce coherence, causing goal drift, where agents gradually move away from the original goals and lose alignment with each other over time~\cite{rath2026agentdriftquantifyingbehavioral}. Existing frameworks also often lack mechanisms for effective context propagation, meaning that agent behavior is not sufficiently conditioned on evolving task state or shared memory. This limitation undermines coordination and shared understanding across agents in complex multi‑agent workflows~\cite{krishnan2025advancingmultiagentsystemsmodel}.
As a consequence, these systems face a trade-off between accuracy and efficiency, constraining their scalability and practical deployment in real-world software engineering tasks ~\cite{han2024llmmultiagentsystemschallenges}.

To address these limitations, we propose ASAD, an adaptive agentic system for debugging that dynamically constructs its team based on the nature and complexity of the faulty code. Rather than relying on a fixed configuration, ASAD begins with an analysis phase in which a central coordinator (the main agent) evaluates the problem’s characteristics and complexity. For simple bugs, it applies a fast, single-pass repair, while for more complex issues, it assembles a tailored team of specialized agents, assigns roles, and orchestrates their collaboration. ASAD integrates three complementary reasoning mechanisms: planning, where the coordinator decomposes debugging into structured subtasks~\cite{huang2024understandingplanningllmagents}; iterative reflection, where agents refine patches based on validation feedback~\cite{madaan2023selfrefineiterativerefinementselffeedback}; and multi-agent collaboration, where agents interact and share information to jointly solve complex debugging tasks~\cite{guo2024largelanguagemodelbased}.

\textcolor{black}{We evaluate ASAD on three established benchmarks: DebugBench~\cite{tian2024DebugBench}, Defects4J~\cite{just2014defects4j}, and CodeFlaws~\cite{Tan2017CodeFlaws}, using six large language models: Llama-3~\cite{grattafiori2024llama3herdmodels}, DeepSeek-V3~\cite{deepseekai2025deepseekv3technicalreport}, Mistral~\cite{jiang2023mistral7b}, Qwen-3~\cite{yang2025qwen3technicalreport}, GPT-5~\cite{singh2025openaigpt5card} and Gemini-1.5~\cite{geminiteam2024gemini15unlockingmultimodal}. The results show that our adaptive system improves bug-fix rates by 12–20\% over chain-of-thought prompting and consistently outperforms static multi-agent systems by 4–9\% in fix precision while reducing average agent usage by 32\%. 
Crucially, our system deploys a single agent for 73\% of syntax and logic errors, and scales up to 4 specialized agents for complex bugs such as concurrency bugs and API misuse patterns. This dynamic scaling delivers higher accuracy without inflating costs, demonstrating that adaptivity, not agent count, is the primary determinant of debugging success.}

This paper extends our previous work~\cite{Majdoub2025Towards} published in the Ideas, Visions and Reflections track of the 33rd ACM International Conference on the Foundations of Software Engineering 2025. The conference paper introduced the concept of adaptive multi-agent orchestration for code debugging and focused on the core idea and initial feasibility. In this work, we enhance the system, expand the evaluation to multiple benchmarks and LLM families, and provide a deeper analysis of system behavior and practical implications. This extension significantly broaden the scope and empirical evidence beyond the original study, demonstrating the effectiveness and robustness of adaptive agent orchestration in automated program repair.

In summary, our contributions are twofold:
\begin{enumerate}
    \item A novel adaptive agentic system for debugging that dynamically configures agent roles, quantity, and collaboration strategy based on real-time analysis of faulty code.
    \item An in-depth empirical evaluation demonstrating the benefits of adaptivity through extensive experiments across three established benchmarks and six LLM families, showing consistent improvements in repair accuracy, computational efficiency, and resource utilization.
\end{enumerate}

The rest of this paper is organized as follows: Section~\ref{sec:background} introduces the core reasoning techniques and the agentic properties of LLM-based systems. Section~\ref{sec:design} presents the architecture and workflow of our approach. Section~\ref{sec:setup} details the experimental setup, and Section~\ref{sec:results} reports the evaluation results. Section~\ref{sec:discussion} provides further analysis, 
Section~\ref{sec:therats} examines threats to validity, Section~\ref{sec:related} surveys related debugging systems, Section~\ref{sec:future} outlines directions for future work
and Section~\ref{sec:conclusion} concludes with a summary of contributions and implications.

\section{Background}
\label{sec:background}

Advances in LLMs have opened up new possibilities in automating software engineering tasks. However, these tasks often require more than language modeling; they require structured reasoning, planning, decision-making, and adaptive behavior. Traditional prompting methods allow LLMs to produce immediate answers, but they struggle with tasks that involve uncertainty, long-term dependencies, or multi-step problem solving. To overcome these limitations, researchers have introduced techniques that enable LLMs to reason deliberately, engage in self-evaluation, utilize tools, and collaborate in specialized roles. These techniques reflect a fundamental shift in the design of LLM-based systems: from static, single-shot generators toward goal-oriented, context-aware, and adaptive agents. This section provides foundational concepts for understanding this transition, focusing first on reasoning mechanisms that enhance LLM problem-solving, and then on the broader notion of agenticness, which frames LLMs as autonomous agents operating in complex environments.

\subsection{Structured Reasoning in LLMs}
\label{sec:reasoning}
Structured reasoning mechanisms, commonly described as agentic design patterns~\cite{agenticDesignP}, extend LLMs beyond single-shot prediction toward deliberate, multi-step problem solving. This section outlines these mechanisms and examines their relevance to automated debugging.

\subsubsection{Reasoning Techniques }
The following paragraphs introduce the key \textcolor{black}{reasoning} techniques: Chain-of-Thought for explicit reasoning, Self-Consistency for improved reliability, Iterative Reflection for self-improvement, Planning for strategic decision-making, Tool Use for grounding in external information, and Multi-Agent collaboration for distributed problem solving.

\subsubsection{Chain-of-Thought}
The most widely used technique is Chain-of-Thought (CoT) prompting~\cite{wei2022chain}. Instead of asking an LLM to output a final answer immediately, CoT prompts the model to articulate its reasoning process explicitly. For example, rather than saying “Fix this bug,” the model is instructed to reason step by step: “First, identify the error type. Next, trace variable values. Then, evaluate possible fixes.” This forces the model to break down complex tasks into manageable subproblems, improving accuracy and making its reasoning traceable.

\subsubsection{Self-Consistency}
To improve consistency and reduce hallucinations, Self-Consistency generates multiple reasoning paths for the same input and selects the most common or best-supported conclusion~\cite{wang2022self}. Rather than relying on a single response, the system produces several candidate solutions, compares them for agreement, and chooses the one that aligns with the majority or satisfies predefined constraints, such as passing all test cases. This simple mechanism significantly increases reliability without requiring additional training data.

\subsubsection{Iterative Reflection}
Even with structured reasoning, initial outputs may still contain subtle errors. Iterative Reflection addresses this by allowing the LLM to evaluate and refine its own work~\cite{madaan2023selfrefineiterativerefinementselffeedback}. The model first generates a draft solution, then critically assesses it for correctness, efficiency, and completeness. Based on this self-evaluation, it revises the output. This cycle can repeat multiple times, mimicking the practice of human developers who review and improve their code before submission. External feedback from another LLM acting as a critic can further enhance quality by identifying logical inconsistencies or style issues.

\subsubsection{Planning}
For more complex tasks, Planning enables the LLM to design a sequence of actions before execution. Rather than jumping to a solution, the model formulates a strategy: “To fix this bug, I must first locate the faulty function, analyze its inputs and outputs, generate candidate fixes, test each one, and validate against the full test suite.” Huang et al.~\cite{huang2024understandingplanningllmagents} identify five key planning strategies:
(1) Task decomposition: splitting a problem into smaller, manageable subtasks;
(2) Multi-plan selection: generating multiple candidate plans and choosing the most promising one;
(3) External module-aided planning: integrating external tools, such as static analyzers or symbolic solvers, to validate or refine plans;
(4) Reflection and refinement: learning from failed attempts to improve future planning decisions;
(5) Memory-augmented planning: storing past experiences or solutions to inform future reasoning, similar to a developer maintaining a personal knowledge base.

\subsubsection{Tool Use}
To ground reasoning in real-world context, Tool Use allows LLMs to interact with external systems. These tools include code executors, web search engines, documentation APIs, and Retrieval-Augmented Generation (RAG) systems that retrieve relevant code examples or library documentation~\cite{chen2025researchlearningreasonsearch}. This capability is essential: no LLM can contain all possible APIs, frameworks, or error patterns. By accessing live, accurate \textcolor{black}{information}, LLMs become far more accurate and trustworthy.
Just as a human developer would consult Stack Overflow or a compiler.

\subsubsection{Multi-Agent Collaboration}
Agentic Designs or Multi-Agent Collaboration distributes reasoning across multiple specialized agents. Each agent is prompted to assume a distinct role, even when all agents use the same underlying model. These agents communicate, share findings, and resolve conflicts, mirroring the structure of human software teams. This division of labor reduces cognitive overload, improves coverage of edge cases, and enhances overall solution quality~\cite{guo2024largelanguagemodelbased,han2024llmmultiagentsystemschallenges}.

While these reasoning techniques have been studied broadly across problem-solving tasks, they are particularly well suited to automated debugging. Debugging inherently involves hypothesis generation, iterative validation, error localization, and decision-making under uncertainty. As a result, techniques such as step-by-step reasoning, reflection, planning, and multi-agent collaboration naturally align with the structure of the debugging process.

\subsubsection{Relevance of Reasoning Techniques for Debugging}
\label{sec:reasoning_debugging}

Debugging is a reasoning-intensive process that involves hypothesis formation, validation under partial observability, and iterative refinement. Unlike code generation, which can often succeed with approximate solutions, debugging demands coherent multi-step reasoning to identify root causes and ensure that fixes are correct and non-regressive, making it particularly sensitive to the limitations of single-shot LLM outputs.
Techniques that encourage step-by-step reasoning, such as Chain-of-Thought, naturally support the decomposition of debugging tasks into interpretable phases including fault localization, state inspection, and fix evaluation~\cite{levin2024chatdbgaipowereddebuggingassistant}. In practice, multiple plausible explanations often exist for the same failure. Approaches based on self-consistency mitigate this ambiguity by exploring alternative reasoning paths and favoring explanations that remain stable across independent analyses, thereby improving robustness in fault diagnosis~\cite{ahmed2023better}.

Debugging is also an iterative activity, where initial fixes may be incomplete or introduce new errors. Reflection-based methods enable LLMs to assess the quality of their own outputs, revise faulty assumptions, and refine patches over successive attempts~\cite{chen2023teachinglargelanguagemodels}. When combined with planning-oriented reasoning, models can adopt a more global debugging strategy, deciding which components to inspect, how to sequence analysis steps, and when to reconsider earlier hypotheses, particularly in large or interdependent codebases.
Effective debugging further requires grounding reasoning in concrete evidence. Tool-augmented approaches allow LLMs to incorporate compiler diagnostics, execution traces, test results, or external documentation into their reasoning process, reducing speculative conclusions and increasing the reliability of generated fixes~\cite{chen2025researchlearningreasonsearch}. Finally, multi-agent reasoning frameworks reflect the collaborative nature of human debugging by distributing analysis across specialized agents, enabling parallel hypothesis generation, cross-verification, and conflict resolution.

Overall, debugging is best viewed as a structured reasoning process rather than a localized code-editing task. Reasoning techniques such as step-by-step decomposition, self-consistency, reflection, planning, tool use, and multi-agent collaboration directly align with the core demands of debugging, making them particularly suitable for LLM-based debugging systems.

\subsection{Agenticness in LLM-Based Systems}
\label{sec:agenticness}

Agenticness describes the degree to which a system can achieve complex goals in complex environments with minimal direct supervision~\cite{shavit2023practices}. In this section, we introduce the core dimensions of agenticness and analyze how they manifest in the context of debugging.

\subsubsection{Core Dimensions of Agenticness}
Shavit et al.~\cite{shavit2023practices} define this capability through four interrelated components: goal complexity, environmental complexity, adaptability, and independent execution.

\subsubsection{Goal complexity} refers to the structure and abstraction level of the objectives an agent must achieve. Highly agentic systems do not merely follow simple commands; they handle vague or underspecified instructions by decomposing them into sequences of interdependent subgoals. This requires hierarchical reasoning and planning, capabilities that align with longstanding work in AI on task decomposition and modular goal representation. Recent advances in LLMs have made this practical: using techniques like Chain-of-Thought prompting~\cite{wei2022chain}, LLMs can now interpret high-level requests and autonomously generate detailed, multi-step plans to fulfill them.

\subsubsection{Environmental complexity} captures the challenges posed by the contexts in which agents operate. Real-world environments are rarely static or fully observable; they may involve dynamic codebases, interactive user inputs, partial information, or concurrent multi-agent interactions. Agentic systems must perceive, interpret, and respond to such uncertainty while maintaining goal-directed behavior~\cite{russell2022human}. This means handling ambiguous inputs, adapting to shifting constraints, and reasoning about consequences in settings where feedback is delayed or incomplete, capabilities that go far beyond simple input-output mapping.

\subsubsection{Adaptability} is the system’s ability to adjust its strategy in response to new information, task variations, or performance feedback. Rather than executing fixed procedures, agentic systems learn from experience, transfer insights across tasks, and refine their behavior on the fly. Modern LLMs exhibit this through mechanisms like in-context learning, where they infer patterns from a few examples without parameter updates; few-shot learning, which enables rapid adaptation to new domains; and meta-learning, where models learn how to learn efficiently across related tasks~\cite{finn2017modelagnosticmetalearningfastadaptation}. These capabilities allow agents to respond intelligently to novel debugging scenarios without retraining.

\subsubsection{Independent execution} is the hallmark of true agency: the capacity to pursue goals autonomously, without step-by-step human intervention. This includes monitoring progress toward a goal, detecting failures, generating new subgoals as needed, and revising plans when obstacles arise. Systems such as AutoGPT~\cite{richter2023autogpt} demonstrate this by chaining together reasoning, tool use, and self-correction over multiple steps to complete open-ended tasks. In the context of program repair, independent execution means an agent can start from a failing test, diagnose the root cause, propose and validate fixes, and iterate, all without external guidance.

\subsubsection{Agenticness of Debugging Systems}
When evaluated through the lens of agenticness~\cite{shavit2023practices}, current LLM-based debugging systems exhibit uneven capabilities across its four dimensions. Most work strongly addresses goal complexity, using advanced reasoning to solve increasingly sophisticated debugging tasks. Some systems also engage with environmental complexity, particularly those that integrate IDE tools, debuggers, or repository-wide context to ground their reasoning in real-world software structures. However, adaptability and independent execution remain largely underdeveloped. Nearly all existing approaches apply the same static workflow to every bug, regardless of difficulty. This leads to two problems: first, simple bugs such as syntax errors are processed with excessive overhead, incurring unnecessary computational cost; second, for complex bugs, fixed agent configurations may lack the right mix of specialists, limiting repair quality. Moreover, in multi-agent settings, static designs often cause agents to drift from the core objective, especially during extended interactions, which can degrade performance unpredictably.
In summary, most current systems operate as reactive tools: they target narrowly defined repair objectives, assume a fixed debugging environment, follow predetermined workflows, and require substantial external control or manual intervention. 
As a result, they exhibit limited agenticness when confronted with complex bugs that demand adaptive reasoning and flexible coordination. 
This motivates the need for debugging systems that can interpret high-level repair goals, respond to diverse program contexts, adapt their internal structure to task difficulty, and execute repairs with minimal supervision.
\newline\newline
To address the challenges of debugging with varying code complexity, we introduce ASAD, an adaptive agentic framework designed to allocate the reasoning and collaboration efforts proportionally to the difficulty of the problem. The approach aims to balance efficiency and effectiveness by using lightweight reasoning for simple bugs while coordinating multiple specialized agents for complex cases. In this section, we start by presenting the overall architecture, then we present its components and the mechanisms that enable adaptive agent deployment and iterative refinement.

\begin{figure*}[t!]
\centering
\color{black}
\includegraphics[width=0.85\linewidth, height=6cm]{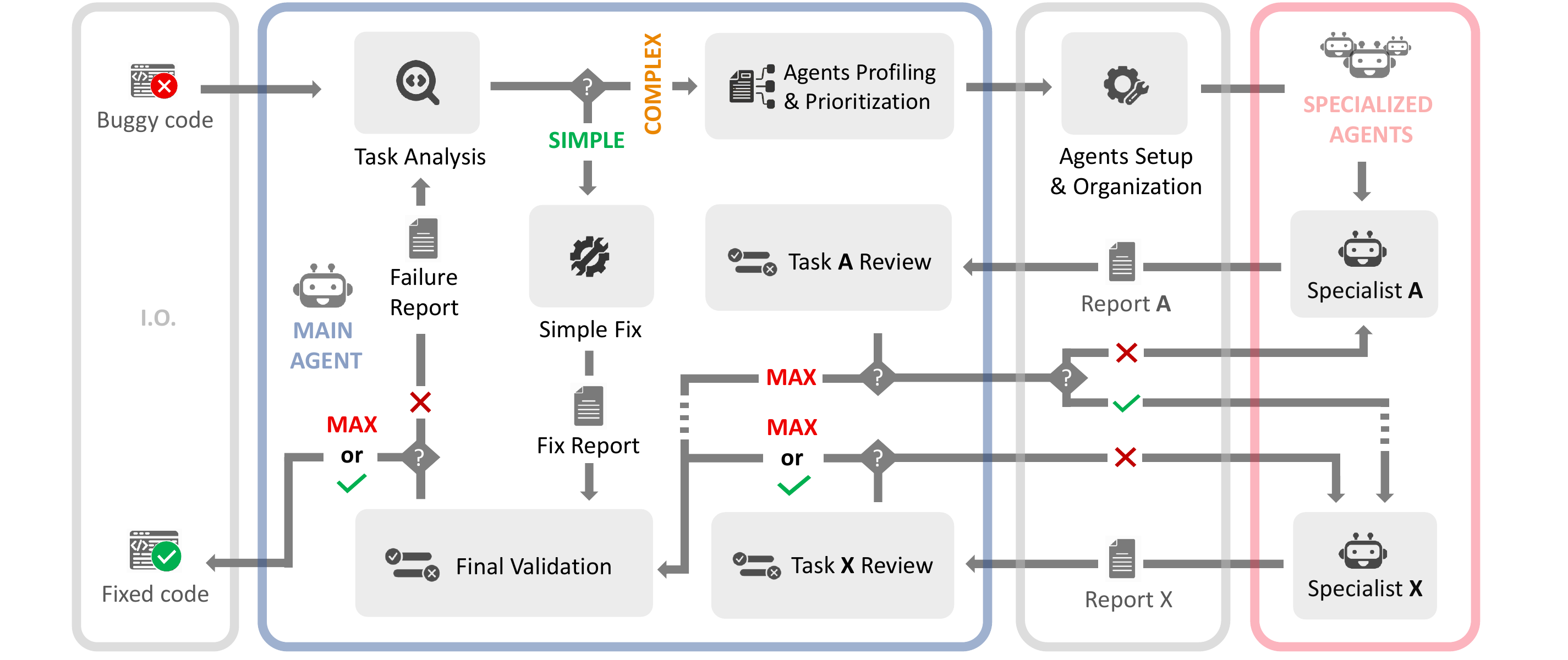}
\caption{Overview of our proposed approach: ASAD}
\label{fig:system}
\end{figure*}

\section{Approach}
\label{sec:design} 
\subsection{Overview of the system}

To be truly adaptive, a system must adjust its strategy based on the characteristics of each problem it encounters. Rather than following a fixed sequence of steps, it should analyze the task, decide how much reasoning is needed, and determine which agents should be involved. We achieve this principle by using the reasoning capabilities of LLMs to assess the buggy code, formulate a plan, and dynamically configure agent roles and interactions. An overview of the system architecture is shown in Figure~\ref{fig:system}.

At its core, our system integrates three complementary reasoning techniques: planning, iterative reflection, and multi-agent collaboration. The system’s behavior depends on the assessed complexity of the bug.
For complex problems, it follows a decomposition-first approach~\cite{huang2024understandingplanningllmagents}. A central main agent first analyzes the bug, decomposes the debugging task into structured subtasks, and designs a coordination plan. It then instantiates specialized agents, assigns each a role, and orchestrates their execution. After agents complete their subtasks, they report their results back to the main agent. Based on this feedback, the main agent may initiate another round of refinement or finalize the solution.
For simple problems, such as minor syntax errors, the system bypasses multi-agent coordination entirely. Instead, the main agent attempts a direct, single-pass repair, avoiding unnecessary overhead. In all cases, the final output undergoes an attentive validation before outputting the solution.
The main agent performs a final check to ensure the correctness of the code and decides whether to go for a new iteration, refining its approach based on the observed failure.
In the following sections, we detail the role, tasks and behaviors of the main agent and the specialized agents it can deploy.

\subsection{Main Agent / Team Leader}

The main agent is the central decision making core of the system. It is responsible for
managing the flow of the task by dynamically guiding the creation
and prioritization of the specialized agents, and the validation of
the results. The main agent is also responsible for integrating feedback from
all agents, adjusting strategies, and guiding the system towards the
efficient solution. 
In the next paragraphs, we detail the tasks of the
main agent.

\subsubsection{Code Analysis}
\label{sec:code analysis}
The process begins with a structured code analysis phase, where the main agent identifies and categorizes defects in the input program. To reduce reliance on purely subjective judgment, we augment the agent’s input with a lightweight static analysis report that provides objective indicators of program structure and potential debugging difficulty.
\color{black}
This analysis extracts five complementary dimensions that characterize the complexity of the buggy program: (1) \textit{Error Severity}, approximated by the number of failing test cases and if available, the runtime or compilation errors observed during execution; (2) \textit{Source-location difficulty}, measured using the Abstract Syntax Tree (AST) depth of the affected code region and the number of non-commenting lines of code (NLOC) in the relevant function; (3) \textit{Program-understanding difficulty}, captured through structural properties such as cyclomatic complexity and AST branching factor; (4) \textit{System-level operations}, identified by the presence of concurrency constructs or resource-management-related API usage; and (5) \textit{Inter-function dependencies}, quantified via call-graph fan-in and fan-out within the file.

These metrics are not used independently, but are provided jointly to the main agent as structured evidence about the nature of the debugging task. The agent reasons over these signals to estimate whether the problem can be resolved through a lightweight single-step repair or requires a more structured multi-agent decomposition.
If the aggregated evidence indicates low complexity, the system proceeds with a direct repair strategy using a single-pass attempt by the main agent. Otherwise, the task is classified as complex, triggering the generation of a multi-step debugging plan and the instantiation of specialized agents. This decision forms the basis of the adaptive orchestration mechanism that guides the remainder of the debugging process.

\color{black}

\subsubsection{Agents Profiling and Prioritization}

When the main agent determines that a problem requires multi-agent decomposition, it generates a structured set of specialized agent profiles. Each profile defines a well-scoped debugging responsibility and serves as an executable specification for instantiating an agent.
\color{black}
Formally, each agent profile consists of three components: (1) \textit{Role}, which defines the debugging function; (2) \textit{Objective}, which specifies the targeted class of issues to be resolved; and (3) \textit{Execution Constraints}, which define the scope of allowed modifications.
Once profiles are generated, the main agent determines an execution order over the instantiated agents. This prioritization is based on dependency relationships between tasks, ensuring that lower-level structural issues are resolved before higher-level semantic reasoning is applied. For instance, syntax-level corrections must precede logic verification, since semantic reasoning assumes syntactically valid code.
The instantiated agents are executed sequentially according to this dependency-aware ordering, an agent $a_i$ can only be executed if all agents with prerequisite responsibilities have completed their tasks successfully. The controller manages the flow of intermediate results between agents, ensuring that each agent operates on the outputs of preceding steps and that the overall debugging process remains coherent and adaptive.

\color{black}

\subsubsection{Iterative Task Review}
In the specialized agents execution phase, the system does not dispatch all agents at once. Instead, it proceeds step-by-step, invoking each agent in the prioritized order determined during the planning stage. After an agent completes its assigned task, it generates a structured report detailing the changes made, the rationale behind them, and any remaining concerns. Upon receiving the report, the main agent evaluates the outcome of the task: if it determines that the result is satisfactory and no critical issues remain, it proceeds to activate the next agent in the sequence.
And if the main agent identifies unresolved problems or deficiencies based on the report, it instructs the current agent to repeat or refine its work until an acceptable result is achieved or a maximum of three attempts is reached.

This supervision, combined with detailed agent reports, ensures that the debugging process remains both reliable and adaptable. It also allows the main agent to maintain transparency and control over each step of the correction process.
Moreover, the iterative review mechanism minimizes the risk of wasted execution, as the system prevents subsequent agents from acting on an incorrect or incomplete solution. By enforcing quality checks at each step, it ensures that time and computational resources are not spent propagating errors, especially when early agents fail to fully resolve the issue.

\subsubsection{Final Validation}
Once all agents have completed their tasks, or if a simple fix path was followed, the main agent performs a final validation, assessing whether the code is now free of issues. If the code is validated as corrected, the system outputs the corrected version to the user. However, if the issue persists, a failure log is generated (including the remaining issues) and passed back to the analysis step. This triggers a new iteration with a revised strategy, which could alter the agent configuration, prompts, or priorities.
To avoid having the system on the same plan as in the previous iteration,  we  explicitly  prompt the main agent to propose a debugging strategy that is different from the ones in the previous iterations. The process repeats until the issue is fixed or the system reaches a maximum of 5 iterations.

\subsubsection{Simple Fix}

If the analysis deems the problem simple, the main agent attempts a direct fix using its internal reasoning capabilities. This is done through a targeted prompt that instructs the agent to correct the identified issue without invoking additional agents. The prompt typically includes the buggy code, a summary of the detected errors, and a directive to produce a corrected version of the code in a single pass. The output of this fix is then passed directly to the validation phase, where it is assessed for correctness. This strategy avoids unnecessary agent deployment and reduces coordination overhead, streamlining the resolution of straightforward issues.
\subsection{Agents Setup \& Workflow Organization}

Once the execution strategy is determined, ASAD initializes the corresponding agents by loading their profiles, assigning code context, and ordering them according to the dependency structure defined during planning.
\color{black}
To balance adaptability and efficiency, ASAD enforces bounded execution at two levels. At the agent level, each specialized agent is allowed up to three refinement attempts when its output is deemed insufficient by the main agent. At the system level, the entire debugging process is limited to five global iterations, where each iteration corresponds to a full cycle of analysis, planning, execution, and validation.
These constraints are motivated by empirical observations of diminishing returns. Additional refinement beyond three attempts per agent rarely improves repair quality while increasing computational cost. Similarly, most successful repairs occur within five global iterations, after which additional cycles yield marginal improvements.
This design ensures that ASAD remains both computationally tractable and robust across varying levels of debugging complexity.

\color{black}

\subsection{Specialized Agents / Team Members}

Specialized agents are instantiated based on the profiles generated by the main agent after analyzing the problem. Each agent has a specific role to perform a particular debugging task. Rather than executing in parallel, agents operate sequentially under the supervision of the main agent. After completing its task, each agent returns a report summarizing its actions and observations. The main agent reviews this report before deciding whether to proceed to the next agent or request further refinement. This review-driven execution protocol enhances robustness by allowing corrections at each step and helps prevent error propagation throughout the debugging process.

\section{Experimental Setup}
\label{sec:setup}

In this section, we describe the experimental setup used to evaluate the effectiveness of our system. We present our research questions, the used benchmarks and baselines, the evaluation metrics, and implementation details. 

\subsection{Research Questions}
In this study, we aim to answer the following research questions (RQs):

\subsubsection{RQ1: How does our adaptive design impact bug-fixing performance compared to traditional approaches?}
We investigate whether ASAD improves bug-fixing performance relative to traditional prompting and fixed multi-agent approaches. 
This involves analyzing the accuracy of generated fixes and the proportion of correctly resolved bugs across benchmarks.

\color{black}

\subsubsection{RQ2: How effectively does the system adapt its agent configuration across bug types and complexity levels, and what is the computational overhead compared to fixed strategies?}
We evaluate whether the system dynamically adjusts the number and roles of agents based on bug complexity, allocating more resources to harder cases while remaining efficient on simpler ones. We further measure inference time and token usage against fixed multi-agent and single-shot baselines to quantify the performance–cost trade-off.

\subsubsection{RQ3: How does adaptive agent selection influence the trade-off between debugging effectiveness and efficiency, and how do different agents contribute to successful repairs?}
We analyze how adaptive agent selection affects the effectiveness–efficiency trade-off by comparing it against fixed-agent configurations across benchmarks and complexity levels. We also examine individual agent contributions by identifying which positions most often produce the first correct patch and how later agents refine or validate solutions, distinguishing gains from adaptation versus simply increasing agent count.

\color{black}

\subsubsection{RQ4: How does the performance of the system vary across different LLMs and experimental tests?}
We first study how different LLMs affect the system effectiveness and efficiency, evaluating the generalizability of its adaptive design across model architectures. Then, we analyze the consistency of results across repeated tests to assess the robustness of the system under stochastic LLM behavior.

\subsection{Benchmarks}

We conduct our experiments on three widely used debugging benchmarks: 
DebugBench, Defects4J and CodeFlaws
These benchmarks collectively cover multiple programming languages (Python, Java, C, and C++) and diverse bug categories, allowing for a comprehensive evaluation of our system across different settings.

\subsubsection{DebugBench}
This benchmark is designed to evaluate the debugging capabilities of LLMs. \textcolor{black}{It consists of 4,253 instances covering four major bug categories: syntax, reference, logic, and multiple bugs; across C++, Java, and Python. The dataset is constructed by collecting correct code snippets from the LeetCode community and systematically injecting bugs using \textcolor{black}{GPT-4} while maintaining rigorous quality control}. DebugBench addresses limitations in prior debugging benchmarks, such as data leakage, limited dataset scale, and insufficient bug diversity. It enables systematic assessment of LLM debugging performance under zero-shot conditions, highlighting strengths and weaknesses across error types.

\subsubsection{Defects4J} 
One of the most established benchmarks for evaluating fault localization and program repair in Java. \textcolor{black}{As of version 3.0.1, it contains 854 reproducible bugs drawn from 17 widely used open-source Java projects, including JFreeChart, Closure Compiler, Gson, Jackson, and Mockito}. Each bug includes a buggy and fixed version, isolated to ensure that changes are limited to the fix itself. The dataset is accompanied by test suites, ensuring that at least one test case fails on the buggy version and passes on the fixed one. Defects4J supports reproducible research and provides metadata facilitating empirical studies on patch size, bug localization, and fix patterns. On average, patches are small and localized, typically around four lines in a single file.

\subsubsection{CodeFlaws}
A large-scale benchmark for automated program repair in C, \textcolor{black}{consisting of 3,902 real-world bugs collected from programming contest submissions. These bugs are classified into 39 syntactic difference-based defect classes, capturing a wide range of common programming errors.} CodeFlaws facilitates objective comparison of different repair tools and enables the study of relationships between fault types and repair effectiveness. It includes buggy and corrected program versions, extensive test suites, and metadata from competitive programming contexts, making it a valuable resource for evaluating automated debugging and program repair techniques.

\subsection{Evaluation Metrics}

To assess the performance of our system, we focus on three main aspects: the bug fix rate, which measures how often the system successfully repairs buggy programs; adaptability, which captures how the system adjusts its strategy based on bug complexity; and computational overhead, which evaluates the resources and time required to perform repairs. We present each metric in the following paragraphs. 

\subsubsection{Bug Fix Rate}
The bug fix rate is the primary metric used to evaluate the effectiveness of the system. It measures the proportion of buggy instances that are successfully repaired by the system, where a repair is considered successful if the generated patch compiles and passes \emph{all} available test cases in the benchmark suite.
Thus, for DebugBench and CodeFlaws, a fix is counted as correct only when the generated code passes the complete test suite provided by the benchmark. For Defects4J, following standard practice in automated program repair~\cite{just2014defects4j}, we distinguish between two related but distinct notions: \textit{plausible fixes} and \textit{correct fixes}.
\color{black}
\begin{itemize}
\item Plausible Fix: A generated patch that compiles successfully and passes all available regression test cases in the benchmark suite. This indicates that the patch is functionally consistent with the test specifications, but does not guarantee correctness beyond the observed tests.

\item Correct Fix: A plausible fix that is equivalent to the ground-truth developer patch in terms of behavior with respect to the benchmark test suite. In practice, this is operationalized as a patch that passes all test cases and is functionally equivalent to the developer-provided solution, when such reference is available. Otherwise, correctness is determined by the benchmark’s official evaluation protocol when it provides additional validation criteria beyond the public test suite.
\end{itemize}
\color{black}
Reporting both metrics is important: \textit{plausible fixes} indicate how often a system produces candidates that satisfy all available test cases in the benchmark, while \textit{correct fixes} capture true reliability by ensuring semantic equivalence with the developer’s intended solution.

\subsubsection{Adaptability} 
Adaptability is crucial for an agentic system that dynamically assigns roles based on task complexity. We measure how effectively the system tailors its debugging strategy, including the number of agents created and iterations performed, in response to varying levels of complexity. We evaluate adaptability by analyzing trends in agent usage across different types of bugs.

\subsubsection{Computational Overhead}
While adaptability and bug fix rate are important, efficiency must also be considered. We evaluate the computational cost incurred by the adaptive design in comparison to fixed multi-agent or single-shot debugging approaches. This is measured in terms of inference time, token consumption. We track the number of iterations required to reach a solution, as well as the computational resources used by the main and specialized agents.

\subsection{Baselines}

We compare our system against both prompting-based methods and state-of-the-art agentic program repair systems to isolate the benefit of adaptive orchestration from improvements in prompting strategies or underlying model capability. Specifically, we evaluate \textit{Zero-shot}, \textit{Chain-of-Thought (CoT)}, and four recent static agentic repair systems: \textit{ChatRepair}~\cite{Xia_2024}, \textit{FixAgent}~\cite{lee2024unifieddebuggingapproachllmbased}, \textit{AdverIntent-Agent}~\cite{ye2025adverIntent}, and \textit{Repair-Agent}~\cite{bouzenia2024repairagentautonomousllmbasedagent}. These baselines cover strong non-adaptive approaches spanning single-prompt, conversational, and fixed multi-agent paradigms.
\color{black}
We further compare against three internal ablation variants to isolate the effect of adaptive routing in our system. All variants preserve the same multi-agent framework but remove dynamic decision-making: (i) \textit{ASAD-Lite}, which always uses a minimal single-agent configuration; (ii) \textit{ASAD-Fixed}, which generates a fixed set of three specialized agents regardless of bug complexity; and (iii) \textit{ASAD-Max}, which always activates the maximum number of agents.

\color{black}
\subsection{Bug Complexity Characterization}

To assess whether the Main Agent's routing decisions track bug complexity, we independently characterize the complexity of each faulty program. For each bug, we identify the buggy-side region affected by the ground-truth fix and compute two measures on that region: the number of changed lines, representing the repair scope, and Halstead Effort, capturing the complexity of the affected code. We use the predefined easy, medium, and hard instances in DebugBench to establish the corresponding complexity ranges and apply the same characterization consistently across CodeFlaws, and Defects4J.

This characterization is performed entirely offline and is independent of ASAD's routing decisions. The ground-truth fixes are used only for this offline analysis and are never available to ASAD during debugging. The Main Agent receives neither the resulting complexity labels nor any information derived from the ground-truth fixes; it independently selects SIMPLE or COMPLEX using only information available during debugging. We then compare its routing decisions with the independently characterized complexity levels to assess whether its adaptive behavior tracks bug complexity.

\color{black}

\begin{table*}[t]
\centering
\caption{Bug fix rates of ASAD across all benchmarks and programming languages, compared to Zero-shot and CoT prompting, using Llama-3, DeepSeek-V3, Mistral, and \textcolor{black}{Qwen-3}.}
\label{tab:fixrates}
\begin{adjustbox}{max width=\textwidth}
\renewcommand{\arraystretch}{1.5}
\begin{tabular}{
p{1.5cm} p{1.3cm}
c c >{\color{black}}c >{\color{black}}c
c c >{\color{black}}c >{\color{black}}c
c c >{\color{black}}c >{\color{black}}c}
\toprule
\textbf{Benchmark} & \textbf{Language}
& \multicolumn{4}{c}{\textbf{Zero-shot}}
& \multicolumn{4}{c}{\textbf{CoT}}
& \multicolumn{4}{c}{\textbf{ASAD}} \\
\cmidrule(lr){3-6}
\cmidrule(lr){7-10}
\cmidrule(lr){11-14}
&
& Llama-3 & DeepSeek-V3 & Mistral & \textcolor{black}{Qwen-3}
& Llama-3 & DeepSeek-V3 & Mistral & \textcolor{black}{Qwen-3}
& Llama-3 & DeepSeek-V3 & Mistral & \textcolor{black}{Qwen-3} \\
\midrule

\multirow{4}{*}{DebugBench}
& Python
& 58.5\% & 61.2\% & 54.7\% & 59.4\%
& 62.9\% & 66.3\% & 59.8\% & 64.8\%
& \textbf{73.7\%} & \textbf{79.5\%} & \textbf{71.8\%} & \textbf{77.3\%} \\

& C++
& 57.4\% & 57.5\% & 53.8\% & 56.2\%
& 64.5\% & 61.4\% & 60.7\% & 62.8\%
& \textbf{75.6\%} & \textbf{74.8\%} & \textbf{72.4\%} & \textbf{74.1\%} \\

& Java
& 52.1\% & 58.2\% & 49.5\% & 56.7\%
& 59.2\% & 63.7\% & 56.8\% & 62.1\%
& \textbf{73.1\%} & \textbf{78.3\%} & \textbf{70.6\%} & \textbf{76.8\%} \\

& \textbf{Avg.}
& 56.2\% & 58.9\% & 52.7\% & 57.4\%
& 62.2\% & 63.8\% & 59.1\% & 63.2\%
& \textbf{74.2\%} & \textbf{77.9\%} & \textbf{71.6\%} & \textbf{76.4\%} \\
\midrule

CodeFlaws
& C
& 52.4\% & 56.8\% & 50.6\% & 55.1\%
& 59.6\% & 63.5\% & 57.4\% & 61.8\%
& \textbf{77.9\%} & \textbf{81.4\%} & \textbf{74.8\%} & \textbf{79.6\%} \\
\midrule

Defects4J
& Java
& 7.9\% & 10.7\% & 6.8\% & 9.9\%
& 9.3\% & 13.2\% & 8.5\% & 12.1\%
& \textbf{24.7\%} & \textbf{32.5\%} & \textbf{22.1\%} & \textbf{29.8\%} \\
\bottomrule
\end{tabular}
\end{adjustbox}
\end{table*}

\subsection{Implementation}

We built ASAD as a modular Python framework that integrates LLMs inside an adaptive, multi-agent system for debugging. The system consists of a central controller and multiple agent modules, which collaborate to analyze buggy programs, classify their complexity, and generate repair strategies. For open-source models, we used Together.ai. For proprietary models such as GPT-5, we used their official APIs. Agents communicate through structured prompts and return their output in a standardized format. To handle interactions between agents, we used task-oriented, role-based prompts with structured output instructions. These prompts ensure consistent, structured communication between the system components and the language models. \color{black}Listing~\ref{prompt1} presents an example of the prompt structure used for Task Analysis, \textit{buggy\_code} refers to the input program under debugging, i.e., the source code instance containing one or more defects, 
while \textit{complexity\_signal}, is a structured set of automatically computed indicators that summarize key properties of the program  as discussed in section \ref{sec:code analysis}.
\color{black}

\begin{lstlisting}[caption={Task Analysis Prompt},label=prompt1, linewidth=0.5\textwidth]
Role: You are the Main Analysis Agent. Your responsibility
is to perform a systematic static analysis of the given 
program, identify defects that prevent correct execution,
and assess debugging complexity using both code content 
and static analysis signals.
Input:
Buggy code: {buggy_code}
Complexity signal: {complexity_signal}
Instructions:
1. Use both the buggy code and the static analysis report 
to assess the complexity of the debugging task and classify 
it as SIMPLE or COMPLEX based on:
   - Error severity
   - Source-location difficulty 
   - Program-understanding difficulty 
   - System-level operations 
   - Inter-function dependencies 
2. Identify and precisely locate all bugs that may prevent
correct execution of the program. For each bug, specify:
   - Bug type (e.g., logic error, API misuse, ...)
   - Location (function name, line number range, ...)
   - Brief explanation of why it causes failure
3. Produce a structured, step-by-step repair plan 
describing how the identified bugs should be fixed.
Output: Return a JSON object with the following schema:
{
  "complexity": "SIMPLE" or "COMPLEX",
  "bugs": [{ "type":_, "location":_, "description":_}],
  "plan": "..."
}
\end{lstlisting}

\section{Results and Analysis}
\label{sec:results}
This section presents the evaluation results of our adaptive debugging system. We report the findings for each of the four research questions.

\subsection{Impact on Bug-Fixing Performance}

\subsubsection{Performance across benchmarks}
\color{black}
Table~\ref{tab:fixrates} reports the bug-fix rates achieved by ASAD across all benchmarks, programming languages, and LLMs, compared with Zero-shot and Chain-of-Thought (CoT) prompting.

Across all evaluated settings, ASAD consistently outperforms both prompting baselines. The improvements are observed for every benchmark.
On DebugBench, ASAD raises the average bug-fix rate from 52.7\%--58.9\% under Zero-shot prompting and 59.1\%--63.8\% under CoT prompting to 71.6\%--77.9\%. Depending on the underlying LLM, this corresponds to absolute improvements of approximately 17--19 percentage points over Zero-shot and 12--14 percentage points over CoT. Similar gains are observed across Python, C++, and Java, indicating that the adaptive strategy remains effective regardless of programming language.

A similar trend is observed on CodeFlaws. ASAD improves bug-fix rates from 50.6\%--56.8\% with Zero-shot prompting and 57.4\%--63.5\% with CoT prompting to 74.8\%--81.4\%, yielding improvements of approximately 18--25 percentage points over Zero-shot and 16--20 percentage points over CoT. These results demonstrate that adaptive collaboration between specialized agents is also effective for competitive-programming style bugs written in C.

Defects4J represents the most challenging benchmark, containing real-world Java bugs that often require long-range reasoning, semantic understanding, and project-level context. Although the absolute repair rates remain lower than those achieved on DebugBench and CodeFlaws, ASAD substantially improves performance across all backbone models, increasing bug-fix rates from 6.8\%--10.7\% (Zero-shot) and 8.5\%--13.2\% (CoT) to 22.1\%--32.5\%. This corresponds to roughly a threefold improvement over conventional prompting, demonstrating that adaptive agent selection remains effective even on realistic software repair tasks with considerably higher complexity.
\color{black} 
Overall, these results show that ASAD consistently delivers large performance gains over prompt-based baselines across all evaluated benchmarks. The improvements remain stable across four different LLM families, suggesting that the benefits arise primarily from the adaptive reasoning framework rather than the capabilities of any particular backbone model.

\begin{table}[t]
\centering
\caption{Number of generated plausible and correct fixes by ASAD on Defects4J, compared to static agentic debugging systems}
\label{tab:Defects4J}

\begin{adjustbox}{max width=\columnwidth,center}{
\renewcommand{\arraystretch}{1}

\begin{tabular}{lccccc}
\toprule
\textbf{Metric}
& \makecell{ChatRepair}
& \makecell{FixAgent}
& \makecell{AdverIntent-\\Agent}
& \makecell{RepairAgent}
& \makecell{\textbf{ASAD}} \\
\midrule
Plausible & $-$ & \textbf{286} & 154 & 186 & 277 \\
\midrule
Correct & 162 & 197 & 140 & 164 & \textbf{223} \\
\bottomrule
\end{tabular}

}
\end{adjustbox}

\end{table}

\begin{table*}[t]
\centering
\color{black}
\caption{Average used agents, time and tokens by ASAD across benchmarks and bug complexity levels.}
\label{tab:agents_summary}
\begin{adjustbox}{width=\textwidth,center}
\renewcommand{\arraystretch}{1.05}
\begin{tabular}{llc*{12}{c}}
\toprule
\multirow{2}{*}{\textbf{Benchmark}} &
\multirow{2}{*}{\textbf{Complexity}} &
\multirow{2}{*}{\textbf{\# Bugs}}
& \multicolumn{3}{c}{\textbf{ASAD + Llama-3}}
& \multicolumn{3}{c}{\textbf{ASAD + DeepSeek-V3}}
& \multicolumn{3}{c}{\textbf{ASAD + Mistral}}
& \multicolumn{3}{c}{\textbf{ASAD + \textcolor{black}{Qwen-3}}} \\
\cmidrule(lr){4-6}
\cmidrule(lr){7-9}
\cmidrule(lr){10-12}
\cmidrule(lr){13-15}
&
&
& Agents & Time & Tokens
& Agents & Time & Tokens
& Agents & Time & Tokens
& Agents & Time & Tokens \\
\midrule

\multirow{3}{*}{DebugBench}
& Low    & 1390 & 1.32 & 35s & 2.3K & 1.12 & 23s & 1.6K & 1.45 & 42s & 2.8K & 1.18 & 26s & 1.8K \\
& Medium & 1946 & 2.01 & 58s & 3.9K & 1.73 & 39s & 2.9K & 2.18 & 70s & 4.7K & 1.81 & 43s & 3.2K \\
& High   &  917 & 3.14 & 96s & 6.8K & 2.74 & 67s & 5.1K & 3.38 &118s & 8.2K & 2.89 & 74s & 5.8K \\

\midrule

\multirow{2}{*}{CodeFlaws}
& Low    & 2781 & 1.18 & 28s & 1.9K & 1.07 & 20s & 1.4K & 1.31 & 35s & 2.3K & 1.12 & 22s & 1.6K \\
& Medium & 1121 & 1.84 & 49s & 3.4K & 1.61 & 34s & 2.5K & 2.01 & 61s & 4.1K & 1.69 & 38s & 2.8K \\

\midrule

\multirow{3}{*}{Defects4J}
& Low    & \textcolor{black}{165} & 1.41 & 41s & 2.7K & 1.21 & 28s & 1.9K & 1.56 & 49s & 3.2K & 1.28 & 31s & 2.1K \\
& Medium & 277 & 2.36 & 73s & 5.1K & 2.03 & 50s & 3.8K & 2.59 & 88s & 6.2K & 2.15 & 56s & 4.2K \\
& High   & 412 & 3.58 &128s & 8.9K & 3.12 & 91s & 6.9K & 3.87 &152s &10.8K & 3.29 &101s & 7.7K \\

\bottomrule
\end{tabular}
\end{adjustbox}
\end{table*}

\subsubsection{Comparison with agentic baselines}

Because Defects4J is one of the standard benchmarks for evaluating automated program repair systems, we further compare ASAD with recent agentic debugging approaches. Table~\ref{tab:Defects4J} summarizes the numbers of plausible and correct fixes.
ASAD produces 277 plausible fixes and 223 correct fixes, achieving the highest number of correct repairs among all compared systems. Although FixAgent generates slightly more plausible patches (286 versus \textcolor{black}{277}), it produces fewer correct repairs (197 versus \textcolor{black}{223}). This indicates that ASAD is more effective at converging toward semantically correct solutions instead of merely generating compilable or test-passing candidates. Compared with ChatRepair, AdverIntent-Agent, and RepairAgent, ASAD improves the number of correct fixes by 61, 83, and 59 bugs, respectively.
These results suggest that adaptive coordination among specialized agents not only increases repair success, but also improves repair precision by enabling more effective validation and refinement throughout the debugging process.

\begin{tcolorbox}[arc=0pt, colback=gray!5,colframe=black!70,title=Answer to RQ1, top=1mm, bottom=2mm, right=2mm, left=2mm]
ASAD consistently outperforms both Zero-shot and Chain-of-Thought prompting across all benchmarks. It achieves substantial absolute improvements on DebugBench and CodeFlaws while nearly tripling repair performance on the challenging Defects4J benchmark. Compared with existing agentic debugging systems, ASAD also achieves the highest number of correct repairs, demonstrating that adaptive multi-agent collaboration improves both repair effectiveness and precision.
\end{tcolorbox}

\color{black}
\subsection{Adaptive Resource Allocation and Computational Efficiency}

A key objective of ASAD is to allocate computational resources according to bug complexity rather than relying on a fixed multi-agent pipeline. Table~\ref{tab:agents_summary} summarizes the average number of activated agents, execution time, and token consumption across all benchmarks, grouped by bug complexity.

Across all benchmarks, ASAD exhibits a clear monotonic increase in resource allocation as bug complexity grows. On DebugBench, the average number of activated agents increases from approximately 1.1--1.5 for low-complexity bugs to 2.7--3.4 for high-complexity bugs across the evaluated LLMs. A similar trend is observed for execution time, which rises from 23--42s to 67--118s, while average token consumption increases from 1.6--2.8K to 5.1--8.2K tokens. Comparable scaling behavior is observed on CodeFlaws and Defects4J, demonstrating that ASAD consistently allocates additional computational resources only as debugging complexity increases.

\begin{figure}[t]
\centering
\color{black}
\includegraphics[width=\linewidth, height=0.27\linewidth]{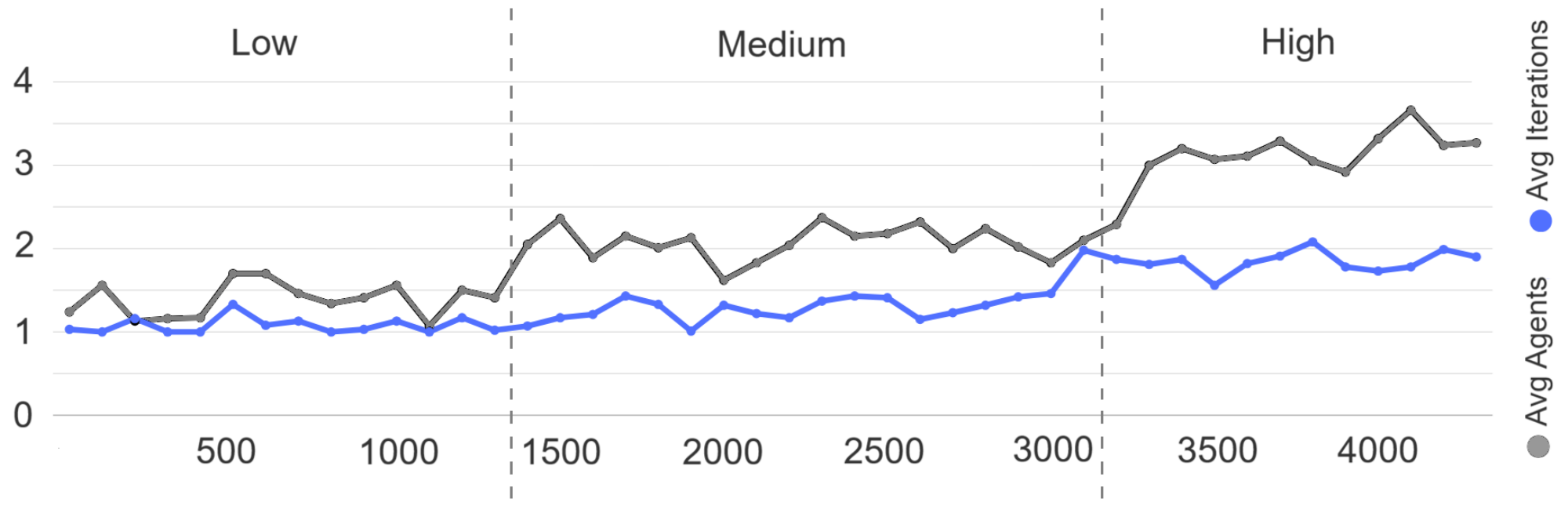}
\caption{Average number of used agents and iterations by ASAD+Deepseek-V3 for each group of 100 DebugBench \textcolor{black}{instances}, sorted by complexity level}
\label{fig:agentsIterations}
\end{figure}

The same adaptive behavior is consistently observed on CodeFlaws and Defects4J. Resource consumption remains modest for simpler bugs but increases progressively for more challenging instances. On Defects4J, the average number of activated agents reaches 3.12--3.87 for the complex bugs. Correspondingly, execution time and token usage also increase, indicating that ASAD invests additional reasoning effort only when the debugging task demands it.
\color{black}
Figure~\ref{fig:agentsIterations} further illustrates this adaptive behavior on DebugBench. As bug complexity increases, ASAD activates more specialized agents and performs additional refinements when necessary, demonstrating that computational resources are allocated progressively rather than remaining fixed throughout the debugging process.
This adaptive allocation provides two important advantages. First, simple bugs are solved with minimal computational cost by activating only a small number of agents, avoiding unnecessary reasoning overhead. Second, more challenging bugs receive additional computational resources, allowing complementary agent roles to collaborate when greater reasoning depth is required. Consequently, ASAD scales its computational effort according to problem difficulty instead of imposing the same execution cost on every debugging task.

\begin{table}[t]
\centering
\caption{Computational costs of ASAD (using DeepSeek-V3) compared to baselines on Defects4J}
\label{tab:overhead}

\begin{adjustbox}{center}{
\renewcommand{\arraystretch}{1}%
\begin{tabular}{lcccc}
\toprule
\textbf{Metric} & Zero-shot & CoT & \textcolor{black}{FixAgent} & \textbf{ASAD} \\
\midrule
Avg Time (s)   & 14.2 & 17.1 & 98.3 & 54.9 \\
\midrule
Avg Tokens (K) & 0.86 & 1.3 & 7.63 & 4.81 \\
\bottomrule
\end{tabular}
}\end{adjustbox}

\end{table}

\begin{table*}[t]
\centering
{
\color{black}
\caption{Ablation Study: Bug fix rate, average used time and tokens by ASAD vs. Fixed-Strategy Baselines across benchmarks and bug complexity levels using DeepSeek-V3}
\label{tab:ablation}
\begin{adjustbox}{width=\textwidth,center}
\renewcommand{\arraystretch}{1.2}
\begin{tabular}{p{1.5cm} p{1.3cm}
cccc
cccc
cccc
cccc}
\toprule
\multirow{2}{*}{\textbf{Benchmark}} &
\multirow{2}{*}{\textbf{Complexity}} &
\multicolumn{3}{c}{\textbf{ASAD-Lite}} &
\multicolumn{3}{c}{\textbf{ASAD-Fixed}} &
\multicolumn{3}{c}{\textbf{ASAD-Max}} &
\multicolumn{3}{c}{\textbf{ASAD}} \\
\cmidrule(lr){3-5}
\cmidrule(lr){6-8}
\cmidrule(lr){9-11}
\cmidrule(lr){12-14}
&
&
\textbf{Fix Rate} & \textbf{Time} & \textbf{Tokens}
& \textbf{Fix Rate} & \textbf{Time} & \textbf{Tokens}
& \textbf{Fix Rate} & \textbf{Time} & \textbf{Tokens}
& \textbf{Fix Rate} & \textbf{Time} & \textbf{Tokens} \\
\midrule

\multirow{3}{*}{DebugBench}
& Low     & 75.2\% & 16s & 1.2K & 83.1\% & 78s & 5.4K & 88.3\% & 105s & 7.6K & 87.1\% & 21s & 1.4K \\
& Medium  & 69.1\% & 21s & 1.5K & 79.4\% & 89s & 6.2K & 77.6\% & 116s & 8.5K & 82.4\% & 36s & 2.6K \\
& High    & 58.9\% & 28s & 2.1K & 69.3\% & 105s & 7.3K & 73.4\% & 128s & 9.6K & 71.3\% & 62s & 4.7K \\
\midrule

\multirow{2}{*}{CodeFlaws}
& Low     & 76.5\% & 15s & 1.1K & 82.4\% & 72s & 4.8K & 84.6\% & 88s & 6.6K & 85.5\% & 18s & 1.2K \\
& Medium  & 63.3\% & 19s & 1.4K & 74.1\% & 80s & 5.3K & 76.4\% & 96s & 7.3K & 78.3\% & 31s & 2.3K \\
\midrule

\multirow{3}{*}{Defects4J}
& Low     & 26.8\% & 17s & 1.0K & 37.4\% & 82s & 5.6K & 34.3\% & 95s & 6.9K & 38.1\% & 25s & 1.6K \\
& Medium  & 19.7\% & 22s & 1.4K & 32.6\% & 90s & 6.4K & 30.5\% & 106s & 7.8K & 31.3\% & 46s & 3.4K \\
& High    & 16.3\% & 29s & 2.0K & 26.4\% & 108s & 7.8K & 25.8\% & 122s & 9.1K & 29.1\% & 85s & 6.3K \\
\bottomrule
\end{tabular}
\end{adjustbox}

}

\end{table*}

To quantify the computational overhead of this adaptive strategy, Table~\ref{tab:overhead} compares ASAD (using DeepSeek-V3) with Zero-shot prompting, Chain-of-Thought prompting, and \textcolor{black}{FixAgent} on Defects4J. As expected, Zero-shot prompting is the fastest and least expensive baseline, followed by Chain-of-Thought prompting, while the fixed multi-agent architecture of \textcolor{black}{FixAgent} incurs the highest computational cost.
ASAD occupies a middle ground between these two extremes. It requires 54.9\,s and 4.81\,K tokens per instance on average, compared with 14.2\,s and 0.86\,K tokens for Zero-shot prompting and 98.3\,s and 7.63\,K tokens for \textcolor{black}{FixAgent}. Thus, although ASAD requires additional computation compared with conventional prompting, it reduces both execution time and token consumption relative to the fixed multi-agent baseline.

Overall, these results demonstrate that ASAD effectively balances computational efficiency and debugging performance. By adapting the number of participating agents to bug complexity, the framework avoids unnecessary computation on simple tasks while allocating additional reasoning capacity when needed, making it more efficient than rigid pipelines.

\begin{tcolorbox}[arc=0pt, colback=gray!5,colframe=black!70,title=Answer to RQ2, top=1mm, bottom=2mm, right=2mm, left=2mm]
ASAD dynamically allocates computational resources according to bug complexity by increasing the number of activated agents, execution time, and token usage only when additional reasoning is needed. Compared with a fixed multi-agent system, ASAD reduces execution time by 44\% and token consumption by 37\% on Defects4J while maintaining superior debugging performance, demonstrating an effective balance between computational efficiency and repair effectiveness.
\end{tcolorbox}

\color{black}

\subsection{Adaptive Agent Selection: Trade-off and Agent Contributions}

\subsubsection{Ablation Study}
To quantify the impact of adaptive orchestration, we compare ASAD with three fixed-strategy baselines: ASAD-Lite, ASAD-Fixed, and ASAD-Max. Table~\ref{tab:ablation} summarizes the results across all benchmarks and complexity levels. ASAD-Lite consistently achieves the lowest computational overhead, with execution times ranging from 15s to 29s and token usage between 1.0K and 2.1K. However, it also yields the weakest debugging performance, with fix rates ranging from 58.9\% to 75.2\% on DebugBench and dropping to 16.3\%--26.8\% on Defects4J. This confirms that relying on a single lightweight agent significantly limits effectiveness on harder debugging tasks.
On the other hand, ASAD-Max achieves competitive or strong fix rates among the baselines, reaching up to \textcolor{black}{73.4}\%--88.3\% on DebugBench and 30.5\%--34.3\% on Defects4J. However, this improvement comes at a substantial computational cost, with execution times up to 128s and token usage reaching 9.6K, representing a multi-fold increase over ASAD-Lite.

ASAD-Fixed provides a stable intermediate baseline, consistently improving over ASAD-Lite in terms of fix rate, but at a higher cost. Across benchmarks, it requires 72s--108s of execution time and 4.8K--7.8K tokens depending on complexity.
In contrast, ASAD achieves a more balanced trade-off between effectiveness and efficiency. On CodeFlaws and Defects4J, it consistently outperforms ASAD-Max and ASAD-Fixed in fix rate (e.g., 78.3\% vs. 76.4\% on CodeFlaws-Medium and 29.1\% vs. 26.4\% on Defects4J-High). On DebugBench, ASAD is slightly below ASAD-Max in fix rate (e.g., 71.3\% vs. 73.4\% on High complexity), but still maintains substantially lower computational cost. Across all settings, ASAD reduces execution time by approximately 50\%--80\% compared to ASAD-Max, depending on benchmark difficulty, while also lowering token consumption by roughly 40\%--70\%.
Overall, the results indicate that performance gains are not simply driven by increasing the number of agents. Instead, adaptively selecting when and how many agents to deploy enables ASAD to maintain strong debugging performance while substantially reducing computational overhead.

\color{black}
\subsubsection{Main Agent routing analysis}
\label{sec:routing_analysis}

\begin{table}[t]
\centering
\renewcommand{\arraystretch}{0.9}
{
\color{black}
\caption{Main Agent classification across independently defined complexity levels, using DeepSeek-V3}
\label{tab:routing_accuracy}
\begin{tabular}{llccc}
\toprule
\multirow{2}{*}{\textbf{Benchmark}} & \multirow{2}{*}{\textbf{Complexity}} &
\multirow{2}{*}{\textbf{\# Bugs}} & \multicolumn{2}{c}{\textbf{Classified as}} \\
\cmidrule(lr){4-5}
& & & \textbf{Simple} & \textbf{Complex} \\
\midrule
\multirow{3}{*}{DebugBench}
& Low    & 1390 & 94.1\% & 5.9\% \\
& Medium & 1946 & 72.4\% & 27.6\% \\
& High   & 917 & 25.8\% & 74.2\% \\
\midrule
\multirow{3}{*}{Defects4J}
& Low    & 165 & 76.4\% & 23.6\% \\
& Medium & 277 & 37.2\% & 62.8\% \\
& High   & 412 & 8.4\%  & 91.6\%\\
\midrule
\multirow{2}{*}{CodeFlaws}
& Low    & 2781 & 97.3\% & 2.7\% \\
& Medium & 1121 & 61.8\% & 38.2\% \\
\bottomrule
\end{tabular}
}
\end{table}

We evaluate the Main Agent's routing decisions against a three-level complexity reference (low, medium, high). Table~\ref{tab:routing_accuracy} reports, for each level, the number of bugs and the proportion classified as Simple or Complex using DeepSeek-V3.
Low-complexity bugs are classified as Simple in 94.1\%
(DebugBench), 76.4\% (Defects4J), and 97.3\% (CodeFlaws) of cases, and high-complexity bugs are classified as Complex in 74.2\% (DebugBench) and
91.6\% (Defects4J) of cases. CodeFlaws contains no high-complexity bugs.
As the Main Agent's output is binary while the reference has three levels, the medium level has no single expected class and is therefore analyzed
through the shift in the classification distribution rather than assigned to either class. 
On all benchmarks, the proportion of Complex classifications
increases monotonically with complexity: 5.9\%, 27.6\%, and 74.2\% on DebugBench; 23.6\%, 62.8\%, and 91.6\% on Defects4J; and 2.7\% and 38.2\% on
CodeFlaws (low and medium only). Medium-complexity bugs consistently occupy an intermediate position between the two extremes, indicating that the Main
Agent's classifications vary systematically with bug complexity.

\color{black}

\begin{table}[t]
\centering
\color{black}
\caption{Agent contribution analysis on 100 randomly sampled DebugBench instances successfully repaired by ASAD using exactly three specialized agents.}
\label{tab:agent_attribution}
\begin{tabular}{lccccc}
\toprule
\textbf{Agent} & \textbf{Pre-Fix} & \textbf{First Fix} & \textbf{Post-Fix} & \textbf{Avg. Time} & \textbf{Avg. Tokens} \\
\midrule
1st  & \textcolor{black}{69\%} & \textcolor{black}{31\%} & -- & 13.2s & 0.87K \\
\midrule
2nd  & \textcolor{black}{31\%} & \textcolor{black}{54\%} & \textcolor{black}{15\%} & 16.1s & 1.1K \\
\midrule
3rd  & -- & \textcolor{black}{15\%} & \textcolor{black}{85\%} & 11.8s & 0.93K \\
\bottomrule
\end{tabular}
\end{table}

\subsubsection{Agent contribution analysis}
To better understand how specialized agents contribute to successful repairs, we conduct a controlled analysis on 100 randomly selected DebugBench instances successfully repaired by ASAD using exactly three specialized agents. For each instance, we record which agent first produces a patch that passes all test cases.
Table~\ref{tab:agent_attribution} summarizes the contribution of each agent position. The second agent is the most effective contributor, acting as the primary fixer in \textcolor{black}{54\%} of successful repairs. This suggests that the first agent often provides an initial but incomplete diagnosis, which the second agent leverages to produce the first correct patch.
The first agent contributes \textcolor{black}{31\%} of successful fixes, typically by producing either partial or occasionally complete solutions early in the process. The third agent accounts for \textcolor{black}{15\%} of first successful fixes, but plays a dominant role in refinement, being responsible for post-fix improvement or validation in \textcolor{black}{85\%} of cases.
\textcolor{black}{These results show that the first two specialized agents account for most first successful fixes, while the third acts primarily as a refinement and verification stage.} The first two specialized agents execute the majority of successful repairs, while the third acts primarily as a refinement and verification stage. These findings \textcolor{black}{support} ASAD’s adaptive design, which prioritizes early specialization and reserves additional reasoning capacity for when it is truly necessary.

\begin{tcolorbox}[arc=0pt, colback=gray!5,colframe=black!70,title=Answer to RQ3, top=1mm, bottom=2mm, right=2mm, left=2mm]
\color{black}
ASAD achieves fix rates comparable to or higher than fixed multi-agent baselines at substantially lower cost, reducing execution time and token usage by approximately 30\%--80\% relative to ASAD-Max. This efficiency stems from adaptive routing: the Main Agent classifies most low-complexity bugs as Simple and most high-complexity bugs as Complex, and the share of Complex classifications increases with complexity on every benchmark. The agent contribution analysis shows that the first two specialized agents produce most first successful fixes, while the third mainly refines and validates them.

\end{tcolorbox}
\color{black}

\subsection{Performance Variability}

\begin{figure*}[t]
\centering
\includegraphics[width=0.8\linewidth]{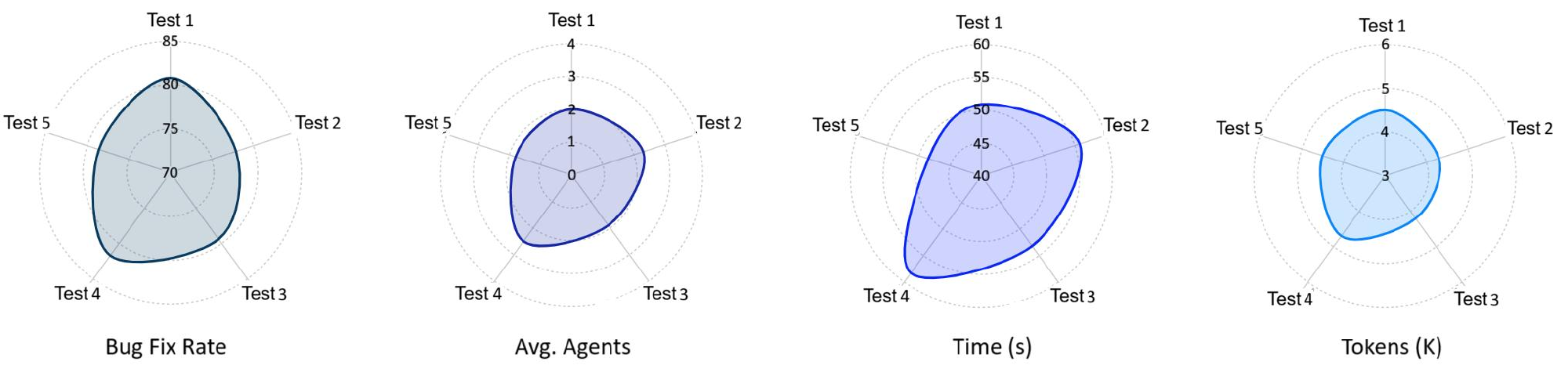}
\caption{Bug fix rate, average number of activated agents, execution time, and token consumption of ASAD over five independent experimental runs (Tests 1--5) using DeepSeek-V3. Each run evaluates the same 100 randomly selected Python instances from DebugBench.}
\label{fig:fiveTests}
\end{figure*}

To assess the robustness of ASAD across different LLM backends, we evaluated the system using six models: Llama-3, DeepSeek-V3, Qwen-3, Mistral, GPT-5, and Gemini-1.5. For each model, we applied the same ASAD pipeline to 500 DebugBench instances. Table~\ref{tab:multimodel} reports the bug-fix rate, average number of activated agents, execution time, and token consumption.
\color{black}

ASAD consistently achieves strong debugging performance across all evaluated models, with bug-fix rates ranging from 68.9\% to 82.2\%. GPT-5 attains the highest repair rate (82.2\%), followed by DeepSeek-V3 (79.8\%) and Gemini-1.5 (76.7\%), while Llama-3, Qwen-3, and Mistral also achieve competitive performance. These results indicate that ASAD generalizes well across diverse LLM families while naturally benefiting from stronger underlying reasoning capabilities.
The adaptive orchestration mechanism also responds differently depending on the backbone model. Qwen-3 requires the fewest agents on average (1.6), whereas Mistral activates the largest number of agents (2.8), suggesting that ASAD allocates additional reasoning resources when the underlying model requires more support. GPT-5 and DeepSeek-V3 achieve the highest repair rates while requiring only 2.1 agents on average, demonstrating that stronger models can solve many bugs with less collaboration.

Execution time follows a similar trend, ranging from 51.1\,s to 68.2\,s. Models activating more agents generally incur higher latency, with Mistral exhibiting both the highest agent usage and the longest execution time. Token consumption remains relatively stable across models (3.6K--4.6K tokens), indicating that ASAD maintains predictable computational overhead despite differences in model architecture and reasoning behavior.

\begin{table}[t]
\centering
\caption{Bug fix rate, average used agents, consumed time and tokens of ASAD across different LLMs on 500 DebugBench instances.}
\label{tab:multimodel}

\begin{adjustbox}{max width=1.2\linewidth,center}{
\renewcommand{\arraystretch}{1.1}%
\begin{tabular}{lcccc}
\toprule
\textbf{Model} & \textbf{Fix Rate} & \textbf{Agents} & \textbf{Time} & \textbf{Tokens} \\
\midrule
Llama-3            & 73.3\% & 2.3 & 55.2s & 4.1K \\
\midrule
DeepSeek-V3        & 79.8\% & 2.1 & 52.8s & 4.6K \\
\midrule
\color{black}Qwen-3      & \color{black}74.2\% & \color{black}1.6 & \color{black}51.1s & \color{black}3.6K\\
\midrule
\color{black}Mistral      & \color{black}68.9\% & \color{black}2.8 & \color{black}68.2s & \color{black}4.3K \\
\midrule
\color{black}GPT-5        & \color{black}82.2\% & \color{black}2.1 & \color{black}51.3s & \color{black}3.7K \\
\midrule
\color{black}Gemini-1.5   & \color{black}76.7\% & \color{black}2.6 & \color{black}61.7s & \color{black}3.9K \\
\bottomrule
\end{tabular}
}\end{adjustbox}
\end{table}

To examine the impact of stochastic generation, we repeated the evaluation with DeepSeek-V3 across five independent runs on a stratified subset of 100 DebugBench instances. As shown in Figure~\ref{fig:fiveTests}, bug-fix rates exhibit limited variation (77\%--82\%, $\sigma = 1.6\%$), with similarly small fluctuations in agent usage, execution time, and token consumption. These results indicate that ASAD maintains stable performance across repeated executions despite the inherent randomness of LLM decoding.
Overall, the results demonstrate that ASAD generalizes effectively across diverse LLM backbones while exhibiting stable and reproducible behavior under stochastic generation.

\begin{tcolorbox}[arc=0pt, colback=gray!5,colframe=black!70,title=Answer to RQ4, top=1mm, bottom=2mm, right=2mm, left=2mm]
ASAD consistently delivers strong debugging performance across diverse LLMs, achieving bug-fix rates between 68.9\% and 82.2\%. Stronger models obtain higher repair rates with relatively few activated agents, while weaker models benefit from additional adaptive coordination. Repeated evaluations with DeepSeek-V3 show only minor stochastic variation ($\sigma = 1.6\%$), confirming that ASAD produces stable and reproducible debugging performance.
\end{tcolorbox}

\color{black}
\section{Discussion}
\label{sec:discussion}

The evaluation of ASAD demonstrates that the adaptive orchestration of LLM-based agents substantially enhances automated debugging performance across diverse benchmarks, programming languages, and model families. Unlike static multi-agent frameworks that enforce rigid pipelines, or simple prompting techniques that lack structured reasoning, ASAD dynamically configures both the number and specialization of agents based on bug complexity. This adaptivity yields massive performance gains—improving bug-fix rates by up to 25 percentage points on synthetic benchmarks and nearly tripling repair performance on the highly complex Defects4J dataset—while simultaneously reducing computational overhead by approximately 44\% compared to fixed multi-agent baselines. These findings provide strong empirical evidence that dynamic, complexity-aware coordination is a critical determinant of efficiency, correctness, and robustness in LLM-based program repair. In this section, we provide a deeper analysis of these results.

\color{black}

\subsection{Interpretation of Results}
Two primary mechanisms explain the observed performance gains. First, \emph{complexity-aware resource allocation} ensures proportional computational utilization. As demonstrated in our resource analysis, ASAD exhibits a clear monotonic increase in agent activation, execution time, and token consumption as bug complexity grows. Trivial syntax or formatting errors are resolved with minimal overhead by activating only a single agent, while complex semantic or logical bugs trigger a cascade of specialized agents. This prevents the wasted computation inherent in fixed pipelines. Second, \emph{structured iterative refinement} allows the system to perform incremental corrections. Our agent attribution analysis reveals a clear division of labor: the first agent typically provides an initial diagnosis or partial fix, the second agent acts as the primary fixer in 54\% of successful repairs, and the third agent predominantly serves as a refiner and validator. This sequential specialization reduces the risk of producing test-suite overfitted patches and allows the system to converge on semantically correct solutions rather than merely compilable ones.

The robustness across different LLM families further highlights the stabilizing effect of the adaptive orchestration layer. Our evaluation across six diverse models shows that ASAD dynamically adjusts its coordination strategy to the underlying model's capabilities: stronger instruction-tuned models (e.g., GPT-5, \textcolor{black}{Qwen-3}, DeepSeek-V3) require fewer agents to achieve high fix rates, while less capable models benefit from increased agent orchestration to compensate for weaker intrinsic reasoning. Furthermore, the low stochastic variance ($\sigma = 1.6\%$) across repeated runs confirms that this structured collaboration yields highly reproducible outcomes despite inherent LLM randomness.

While absolute repair rates on Defects4J remain lower than on synthetic benchmarks due to the profound semantic dependencies and project-level context required for real-world Java bugs, the \emph{relative} improvement is striking. ASAD nearly triples the repair performance of conventional prompting on this dataset. This indicates that while highly complex bugs remain challenging for any automated system, adaptive multi-agent collaboration is crucial when long-range reasoning and deep semantic understanding are demanded by the task.

\subsection{Comparison with Static and Baseline Approaches}
The comparative analysis against both static multi-agent frameworks and standard prompting techniques underscores the absolute necessity of adaptivity. Previous multi-agent debugging systems, such as FixAgent, rely on \emph{static pipelines} where the number, type, and order of agents are fixed. Our results reveal that such rigidity leads to significant inefficiencies. On Defects4J, FixAgent generated more plausible patches than ASAD but fewer correct fixes, indicating a higher incidence of test-suite overfitting and a failure to converge on true semantic correctness. Similarly, our ablation study comparing ASAD against fixed internal variants (\textit{ASAD-Fixed} and \textit{ASAD-Max}) shows that simply increasing the number of agents yields diminishing marginal returns at the cost of massive computational overhead (more than 6$\times$ the token consumption of the lightweight baseline). ASAD, by contrast, matches or exceeds the fix rates of static configurations while reducing execution time and token consumption by 44\%--55\%.

Conversely, standard prompting techniques (Zero-shot and Chain-of-Thought) lack the structured collaboration necessary for complex debugging. While they are computationally cheap, their performance plateaus significantly on challenging benchmarks. ASAD bridges this gap by providing a reasoning-driven coordination mechanism that scales its computational effort only when the task demands it. This selective deployment not only improves efficiency and correctness but also prevents the over-correction and hallucination often seen in unconstrained LLM generation. Ultimately, these insights underline that dynamic, context-sensitive orchestration is vastly superior to both rigid parallelism and unstructured single-agent prompting, establishing adaptivity as the key enabler for next-generation automated debugging systems.

\color{black}
\subsection{Data Contamination in Established Benchmarks}

Data contamination in established benchmarks like Defects4J and CodeFlaws threatens the external validity of our findings, as frontier models may have encountered these datasets during pre-training. Although recent studies identify Defects4J as the benchmark with the highest overall memorization risk, they also confirm that modern models with substantially larger training budgets, specifically the Llama-3 and Mistral families used in our evaluation, exhibit significantly lower leakage risks~\cite{ramos2025largelanguagemodelsmemorizing}. By measuring Negative Log-Likelihood and 5-gram accuracy, researchers demonstrated that these newer models treat established benchmarks similarly to unseen code, avoiding the verbatim memorization observed in older architectures. Our own experimental results corroborate this limited memorization. If the evaluated models had simply retrieved memorized solutions for Defects4J, we would expect near-perfect repair rates. Instead, the maximum fix rate reaches only 32.5\%. This imperfect performance demonstrates that the models must actively reason through complex semantic dependencies to generate valid patches. Furthermore, our adaptive approach consistently improves fix rates over static baselines across all complexity levels. If data contamination drove the performance, this adaptive orchestration would not yield such structured and consistent improvements.

\color{black}

\section{Threats to validity}
\label{sec:therats}

In this section, we examine possible threats to the validity of our empirical findings.

\subsection{Internal Validity}
LLMs exhibit inherent non-determinism due to stochastic decoding strategies and potential API-level variability, which may affect result stability. To mitigate this threat, we conducted five independent replications on a stratified sample of 100 DebugBench instances per model, fixing temperature parameters ($t=0.2$) and random seeds where API-supported. Performance measurements yielded a mean fix rate with a narrow standard deviation $\sigma=1.6\%$ across replications.
Another threat concerns test-suite overfitting in repair validation. Passing benchmark test cases does not guarantee semantic correctness, particularly on Defects4J where limited coverage enables plausible but incorrect patches. We mitigated this by distinguishing plausible from correct fixes using developer-provided oracles, conducting manual inspection on all 223 patches classified as correct.
Subjectivity in complexity assessment also presents another threat, as agent allocation depends on the main agent's internal judgment without objective ground truth. We addressed this by formalizing complexity criteria into five measurable dimensions (Error Severity, Source-location difficulty, Program-understanding difficulty, System-level operations, and Inter-function dependencies.) with explicit decision thresholds, validating assessments against human judgments on 100 bugs.

\subsection{External \textcolor{black}{Validity}}

The applicability of the proposed approach to other datasets constitutes an external validity threat. Results observed on a given benchmark may not generalize uniformly to other debugging datasets with different characteristics. To limit this threat, we evaluated ASAD on multiple benchmarks (DebugBench, Defects4J, and CodeFlaws) covering different programming languages, bug types, and dataset properties, thereby reducing reliance on any single benchmark. 
Nevertheless, the benchmarks considered in this study primarily focus on single-file or unit-level defects and do not fully capture the complexity of large industrial software systems. As a result, the effectiveness of ASAD in repository-scale settings involving multi-module dependencies, build systems, and CI/CD pipelines remains uncertain. Evaluating the approach on repository-level benchmarks and in real-world industrial environments is an important direction for future work.
Generalization across LLMs represents another external threat. The evaluation included several model families available in 2025, including GPT-5, DeepSeek-V3, Gemini-1.5, and Llama-3. While newer models may behave differently, consistent performance trends were observed across the evaluated models, suggesting that the findings are not tied to a specific LLM.

\section{Related Work}
\label{sec:related}

In this section, we review recent research on the use of LLMs for debugging, focusing on reasoning techniques as well as agent-based approaches.

\subsection{LLM Reasoning Techniques in Debugging}

Recent work has explored a range of reasoning techniques to improve the debugging capabilities of LLMs. Levin et al.~\cite{levin2024chatdbgaipowereddebuggingassistant} introduce ChatDBG, a debugging assistant that applies Chain-of-Thought prompting to generate step-by-step explanations for its proposed fixes, making the reasoning process transparent and verifiable. Chen et al.~\cite{chen2023teachinglargelanguagemodels} propose “Self-Debugging,” in which LLMs iteratively refine code based on actual execution outcomes, using runtime feedback to guide corrections. Ahmed et al.~\cite{ahmed2023better} enhance patch generation through Self-Consistency, producing multiple candidate fixes and selecting the most coherent or frequently occurring solution.
Other approaches integrate reasoning with structured workflows. Kang et al.~\cite{kang2023explainableautomateddebugginglarge} present AutoSD, a system that formulates debugging as scientific inquiry: the LLM generates hypotheses about the bug’s root cause and then uses a debugger to test them. Similarly, Lee et al.~\cite{lee2024unifieddebuggingapproachllmbased} develop FixAgent, which combines multi-agent collaboration and tool use to mimic real developer workflows, separating diagnosis, patching, and validation into distinct phases.
To further extend reasoning, recent systems embed retrieval, planning, and tool interaction directly into the inference loop. Chen et al.~\cite{chen2025researchlearningreasonsearch} integrate retrieval into multi-step reasoning, allowing LLMs to dynamically fetch and incorporate external knowledge during problem solving. Yan et al.~\cite{yan2024mirrormultipleperspectiveselfreflectionmethod} combine reflection with tool execution, using both intra-agent self-critique and inter-agent feedback to revise plans before and after invoking tools. Finally, Schick et al.~\cite{schick2023toolformerlanguagemodelsteach} demonstrate that tool invocation can be learned and embedded directly into the LLM’s generation process, enabling the model to reason through executing code, observing outcomes, and updating its plan in response.

\subsection{LLM-based Agentic Systems for Debugging}

Building on these reasoning patterns, recent research has developed multi-agent systems that distribute debugging responsibilities across specialized LLM agents. Qin et al.~\cite{qin2024agentflscalingllmbasedfault} propose AgentFL, a modular framework for bug localization that employs four distinct agents (Test Code Reviewer, Source Code Reviewer, Software Architect, and Software Test Engineer) to analyze different aspects of the codebase using test and documentation signals. Kang et al.~\cite{kang2023explainableautomateddebugginglarge} further enhance interpretability by prompting agents to engage in hypothesis-driven reasoning, requiring them to justify and test each step before proposing a fix.
More advanced systems coordinate agents across larger codebases. Similarly, Alibaba’s LingmaAgent~\cite{ma2025alibabalingmaagentimprovingautomated} integrates LLM agents directly into the IDE, allowing them to gather contextual information from the entire codebase to resolve issues more accurately.
Structured coordination is also central to MAGIS~\cite{tao2024magisllmbasedmultiagentframework}, which assigns roles such as manager, developer, and quality assurance agent to enforce task decomposition, accountability, and verification.

\section{Future Work}
\label{sec:future}

We will scale ASAD to repository-level debugging, where agents must coordinate across dozens of files while respecting import dependencies, build constraints, and cross-module data flows. This requires capabilities for traversing call graphs, propagating fixes across coupled components, and validating changes against project-wide test suites.
We will also close the validation gap by integrating runtime feedback into agent reasoning loops. Agents will execute proposed patches against dynamically generated tests targeting edge cases absent from benchmark suites, using observed crashes, timeouts, and silent failures to detect overfitting and refine patches.
We further plan to connect ASAD to developer ecosystems through static analyzers (e.g., Coverity, SonarQube) for fault pre-filtering, IDE linters for project-specific conventions, and version control systems for assessing patch impact through change histories and reviewer feedback.
Finally, we will study human-in-the-loop debugging, measuring how agent explanations of fault hypotheses and repair strategies affect developer trust, intervention timing, and collaborative efficiency.

\section{Conclusion}
\label{sec:conclusion}

Static multi-agent frameworks deploy fixed teams irrespective of bug complexity, leading to inefficient resource use on simple syntax errors and insufficient capacity for complex, multi-file faults. ASAD mitigates both issues by constructing its agent team after analyzing the fault. In our experiments, a single repair agent successfully resolves 73\% of straightforward bugs, while more complex cases dynamically trigger a team of specialized agents.
Our evaluation on DebugBench, Defects4J, and CodeFlaws substantiates the effectiveness of this approach. ASAD improves fix accuracy by 12–20\% over chain-of-thought prompting and by 4–9\% over static multi-agent systems. These gains were achieved with 32\% lower average agent usage, as ASAD scales down for simple bugs instead of imposing uniform multi-agent overhead.
Together, these results suggest a broader design principle for agentic software engineering: agent configuration should be a function of problem complexity rather than a fixed predefined choice.

\section*{Data Availability}

All datasets used in this study are publicly available. We also release the implementation, prompts, and outputs at~\cite{asad_repo}.



\bibliographystyle{IEEEtran}
\bibliography{ASAD}

\end{document}